\documentclass[twocolumn,10pt,superscriptaddress,amsmath,amssymb,aps,pra,floatfix,]{revtex4-2}
\usepackage{graphicx}
\usepackage{bm}
\usepackage{times}
\usepackage{amsmath,amssymb}
\usepackage{mathtools}
\usepackage{enumerate}
\usepackage{multirow}
\usepackage{xcolor}
\usepackage{braket}

\usepackage[colorlinks=true,linkcolor=blue,citecolor=blue, linktocpage=true,breaklinks=true]{hyperref}

\newcommand{\eq}{\begin{equation}}
\newcommand{\en}{\end{equation}}
\newcommand{\eqa}{\begin{eqnarray}}
\newcommand{\ena}{\end{eqnarray}}
\newcommand{\tr}{\mathrm{Tr}}

\newcommand{\n}{\nonumber\\}

\begin{document}

\title{Emergent mirror symmetry in the optimization of the central-spin quantum battery}

\author{Hui-Yu Yang}
\affiliation{School of Physics, Northwest University, Xi'an 710127, China}

\author{Kun Zhang}
\email{kunzhang@nwu.edu.cn}
\affiliation{School of Physics, Northwest University, Xi'an 710127, China}
\affiliation{Shaanxi Key Laboratory for Theoretical Physics Frontiers, Xi'an 710127, China}
\affiliation{Peng Huanwu Center for Fundamental Theory, Xi'an 710127, China}
\affiliation{Fundamental Discipline Research Center for  Quantum Science and Technology of Shaanxi Province, Xi'an 710127, China}

\author{Xiao-Hui Wang}
\email{xhwang@nwu.edu.cn}
\affiliation{School of Physics, Northwest University, Xi'an 710127, China}
\affiliation{Shaanxi Key Laboratory for Theoretical Physics Frontiers, Xi'an 710127, China}
\affiliation{Peng Huanwu Center for Fundamental Theory, Xi'an 710127, China}
\affiliation{Fundamental Discipline Research Center for  Quantum Science and Technology of Shaanxi Province, Xi'an 710127, China}

\author{Hai-Long Shi}
\email{hailong.shi@ino.cnr.it} 
\affiliation{INO-CNR and LENS, Largo Enrico Fermi 2, 50125 Firenze, Italy}

\begin{abstract}
Quantum batteries provide a useful setting for exploring nonequilibrium many-body effects in energy storage. 
Here we investigate the optimization of a quantum battery based on the central-spin model. 
We identify two complementary structural indicators associated with the effective charging dynamics: one yields an upper bound on the average charging power, while the other characterizes the buildup of stored energy. 
We show that these two indicators are jointly optimized at a distinguished initial charger excitation number, which selects a particular Dicke sector of the model. 
At this common optimal point, the effective charging Hamiltonian becomes exactly mirror symmetric, suggesting mirror symmetry as a useful structural indicator for optimizing the performance of quantum batteries in terms of both charging power and energy storage. 
We further show that the corresponding optimal dynamics can be closely approximated by product initial states, in particular by spin coherent states whose excitation-number distribution is centered at the symmetry-selected point. 
Our results establish a direct connection between charging performance, optimal-state structure, and emergent symmetry in the central-spin quantum battery, and suggest symmetry as a useful organizing principle for efficient charging in interacting many-body quantum systems.
\end{abstract}

\maketitle
\section{Introduction}

Quantum batteries (QBs) have emerged as a useful framework for investigating how genuinely quantum features affect energy storage and transfer in microscopic systems~\cite{PhysRevE.87.042123,campaioli2024colloquium,hymas2026superextensive,andolina2018charger,farina2019charger}. 
Beyond the basic task of storing energy, current research has increasingly focused on the ultimate performance of QBs, including the achievable charging power~\cite{campaioli2017enhancing,ferraro2018high,binder2015quantacell,zakavati2021bounds,moraes2021charging,pokhrel2025large}, the maximum extractable~\cite{garcia2020fluctuations,andolina2019extractable,shi2022entanglement,song2024evaluating,bhattacharyya2024noncompletely} or storable energy~\cite{arrachea2023energy,grazi2024controlling,yang2023battery,caravelli2021energy,crescente2020charging}, and the possible advantages arising from collective many-body effects~\cite{gyhm2024beneficial,andolina2025genuine,gyhm2022quantum,rossini2020quantum,divi2025sachdev,shi2025quantumchargingadvantagemultipartite}. 
In this context, the interplay among interaction geometry, state preparation, and nonequilibrium dynamics has become a central theme in the study of quantum energy devices~\cite{joshi2022experimental,rodriguez2024optimal,andolina2019quantum}.

Among the various many-body platforms, the central-spin model provides a minimal but nontrivial setting for exploring these questions~\cite{yang2025optimal,liu2021entanglement,xu2026non,peng2021lower}. 
In the setup considered here, $N_b$ central spins form the battery cells and are collectively coupled to $N_c$ bath spins that act as charging units. 
This architecture captures collective energy exchange in a simple many-body setting while retaining essential features of the charging dynamics. 
It therefore offers a natural platform for studying the optimization of QBs, in particular the relation between performance bounds and the structure of the corresponding optimal states~\cite{tiwari2023quantum,bortz2007exact}. 
Moreover, the central-spin model has long served as a paradigmatic framework for quantitatively understanding decoherence in nitrogen-vacancy centers in diamond and for exploring entanglement dynamics in interacting low-dimensional quantum systems~\cite{li2020dynamics,fan2023collapse,childress2006coherent,hanson2008coherent}.

A major goal in QB research is to identify fundamental constraints on performance and to understand the structure of the states that optimize them. 
Although a variety of bounds on charging power and energetic performance have been established in different settings~\cite{julia2020bounds,pg2026upper,mohan2026fundamentallimitationsreliabilitiespower,shukla2025versuschargerperformanceoptimization}, much less is known about how such bounds are connected to the internal structure and symmetry of optimal states in concrete many-body models. 
In particular, for the central-spin QB, it remains unclear whether the optimization of charging power and energy storage selects special classes of initial states with distinctive symmetry properties. 
Previous studies have shown that an emergent SU(2) symmetry can ensure optimal energy storage in central-spin and Tavis-Cummings QBs, but these results typically rely on limits involving substantial physical resources, such as large photon numbers~\cite{yang2024optimal,yang2025optimal}. 
It is therefore desirable to clarify, in a more general optimization setting, whether symmetry can emerge directly from the optimization problem itself and serve as a useful structural indicator for designing efficient many-body charging dynamics.

In this work, we study the optimization of a central-spin quantum battery through the structure of its effective charging Hamiltonian. 
Rather than treating charging power and energy storage separately, we identify two complementary structural indicators associated with the hopping profile in the invariant excitation subspace. 
The first yields an upper bound on the average charging power, while the second characterizes the buildup of stored energy through the fully charged channel. 
By analyzing their dependence on the initial charger excitation number, we determine a distinguished Dicke sector in which the two indicators are jointly optimized. 
At the same point, the effective hopping profile becomes exactly mirror symmetric, revealing a direct connection between charging optimality and an emergent symmetry of the model. 
Finally, we show that the resulting optimal dynamics can be closely approximated by simple product initial states, especially by spin coherent states whose excitation-number distribution is centered at the symmetry-selected point.

The remainder of this paper is organized as follows. 
In Sec.~\ref{sec:model}, we introduce the central-spin QB and define the relevant performance quantities. 
In Sec.~\ref{sec:bound}, we introduce two complementary structural indicators for the average charging power and the buildup of stored energy. 
In Sec.~\ref{sec:mirror}, we analyze the common optimal point of these two indicators, show how it gives rise to emergent mirror symmetry in the effective charging Hamiltonian, and discuss the resulting dynamical consequences. 
In Sec.~\ref{sec:simulate}, we discuss how the optimal charging dynamics can be approximated using product initial states. 
Finally, Sec.~\ref{sec:conclusion} presents our conclusions.

\begin{figure}[t]
\centering
\includegraphics[width=1\linewidth]{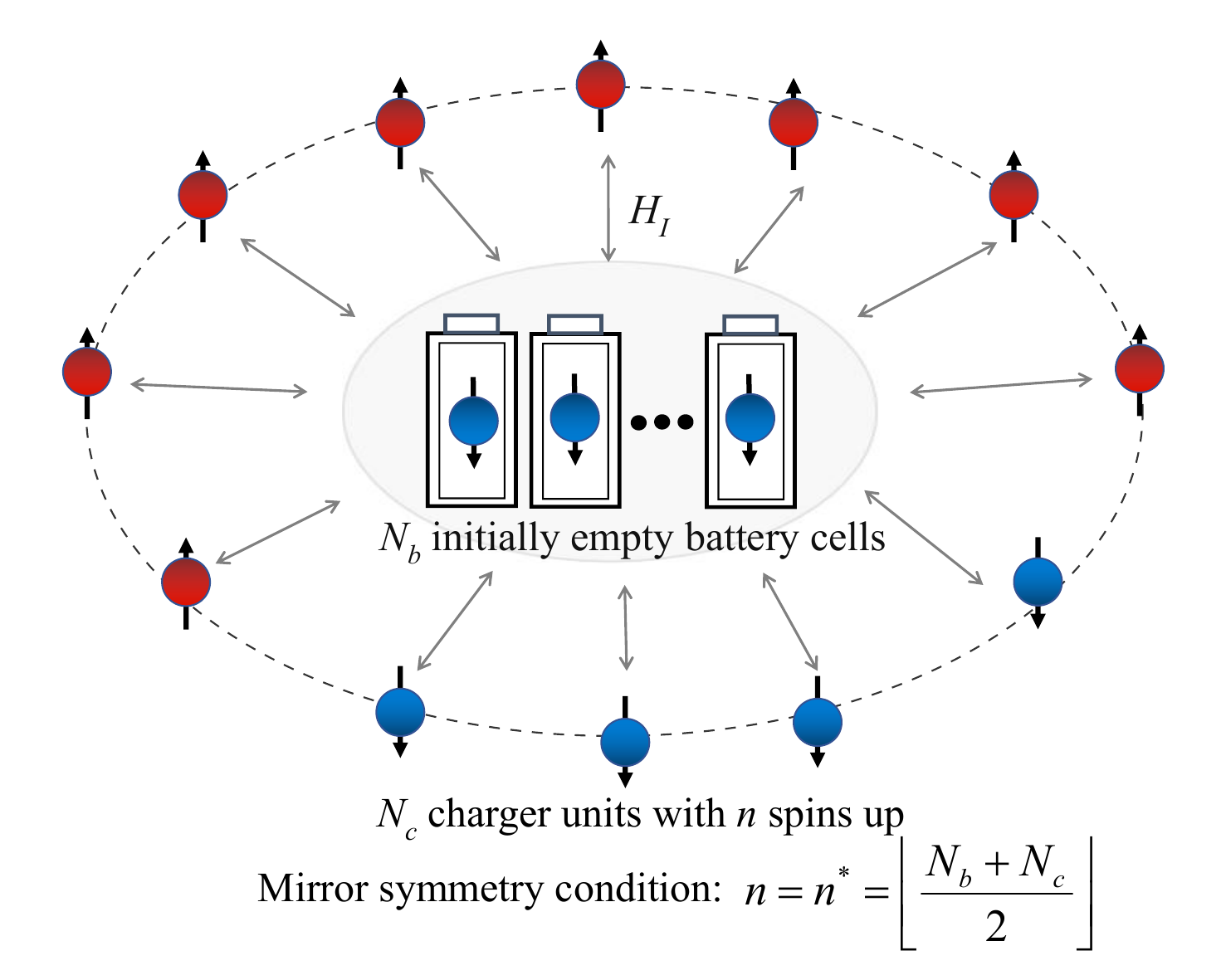}
\caption{
Schematic illustration of the central-spin quantum battery. 
The battery consists of \(N_b\) central spins, initially prepared in the empty state, and is collectively coupled through the exchange interaction \(H_I\) to a charger composed of \(N_c\) spins. 
The charger is initially prepared in a Dicke state with \(n\) spin excitations. 
The optimization of the effective charging dynamics selects the symmetry point
\(n=n^\ast=\lfloor (N_b+N_c)/2\rfloor\), at which the hopping profile in the invariant excitation subspace becomes mirror symmetric.
}
\label{fig:CSM}
\end{figure}

\section{Central-Spin Quantum Battery}
\label{sec:model}

We consider a central-spin QB in which $N_b$ central spins form the battery cells and are collectively coupled to a charger composed of $N_c$ spins, see Fig.~\ref{fig:CSM}.
The total Hamiltonian is
\begin{align}\label{total_H}
H &= H_b + H_c + H_I , \n
H_b &= \omega_b S_z, \qquad H_c = \omega_c J_z, \n
H_I &= A \left( S_{+} J_{-} + S_{-} J_{+} \right),
\end{align}
where $H_b$ and $H_c$ describe the battery and charger, respectively, and $H_I$ accounts for the collective exchange interaction between them. 
Here $S_\alpha=\frac{1}{2}\sum_{j=1}^{N_b}\sigma_j^\alpha$ and $J_\alpha=\frac{1}{2}\sum_{k=1}^{N_c}\sigma_k^\alpha$ ($\alpha=x,y,z$) are the collective spin operators of the battery and charger, with $\sigma_j^\alpha$ and $\sigma_k^\alpha$ the Pauli operators acting on the $j$th battery spin and the $k$th charger spin, respectively. 
The corresponding raising and lowering operators are $S_\pm=S_x\pm iS_y$ and $J_\pm=J_x\pm iJ_y$. 
Physically, the interaction term $A(S_+J_-+S_-J_+)$ describes coherent exchange of excitations between the battery and the charger.

To interpret this model as a QB, we regard the battery spins as the subsystem in which energy is stored, while the charger provides the energy input through the interaction term. 
To ensure that the energy stored in the battery originates solely from the charger, we impose the resonance condition $\omega_b=\omega_c\equiv\omega>0$, under which $[H_b+H_c,H]=0$.

Initially, the battery is prepared in its ground state $|0\rangle_b=|\downarrow_1,\downarrow_2,\ldots,\downarrow_{N_b}\rangle$ of $H_b$, while the charger is initialized in a Dicke state with $n$ excitations,
\begin{equation}
|n\rangle_c
=
\frac{1}{\sqrt{\binom{N_c}{n}}}
\sum_k
O_k
\Big(
|\uparrow_1,\ldots,\uparrow_n,\downarrow_{n+1},\ldots,\downarrow_{N_c}\rangle
\Big),
\end{equation}
where $O_k$ denotes all distinct permutations. 
Because the total excitation number is conserved, the dynamics starting from
\begin{equation}\label{initial_state}
|\psi(0)\rangle = |0\rangle_b\otimes |n\rangle_c
\end{equation}
remains in the invariant subspace
\begin{equation}\label{invariant_subspace}
\mathcal H_n
=
\mathrm{span}
\left\{
|j\rangle_b\otimes|n-j\rangle_c
\;\middle|\;
j=0,1,\ldots,d
\right\},
\end{equation}
with $d=\min\{N_b,n\}$. 
Here $|j\rangle_b$ and $|n-j\rangle_c$ denote Dicke states of the battery and charger, respectively. 
When $n\ge N_b$, the charger carries enough excitations to fully charge the battery, so that the fully charged state $|N_b\rangle_b$ is accessible within the same invariant sector. 
This is the regime considered in the following.

Within $\mathcal H_n$, the Hamiltonian $H$ in Eq.~\eqref{total_H} is represented by the $(d+1)\times(d+1)$ tridiagonal matrix
\begin{equation}\label{eq:tridiag}
\bm{H}\equiv H|_{\mathcal H_n}=
\begin{pmatrix}
b_0 & u_1 & 0 & \cdots & 0 \\
u_1 & b_1 & u_2 & \cdots & 0 \\
0 & u_2 & b_2 & \cdots & 0 \\
\vdots & \vdots & \vdots & \ddots & u_d \\
0 & 0 & 0 & u_d & b_d
\end{pmatrix}.
\end{equation}
Here and in what follows, bold italic symbols denote operators restricted to the invariant subspace $\mathcal H_n$. 
The matrix elements are
\begin{align}\label{off_diagonal}
u_j(n) &= A\sqrt{j(N_b-j+1)(N_c-n+j)(n-j+1)}, \n
b_j(n) &= \omega\left(n-\frac{N_b+N_c}{2}\right).
\end{align}
Since $b_j(n)$ is independent of $j$, the nontrivial structure of the charging dynamics is entirely encoded in the hopping amplitudes $u_j(n)$, up to an overall energy shift.

Assuming that $\bm H$ is diagonalized by a unitary matrix $\bm U$ such that $\bm H=\bm U\bm E\bm U^\dagger$, the state vector in the invariant subspace is
\begin{equation}\label{eq:WaveFunc}
\bm{\psi}(t)=\bm{U} e^{-i\bm Et}\bm U^\dag (1\ 0\ \ldots\ 0)^T,
\end{equation}
where $\bm{\psi}(t)=\bigl(\psi_1(t),\psi_2(t),\ldots,\psi_{d+1}(t)\bigr)^T$ denotes the vector of time-dependent amplitudes in the basis $\mathcal H_n$. 
Accordingly, the corresponding evolved state reads
\begin{equation}\label{eq:total-state}
|\psi(t)\rangle
=
\psi_1(t)|0\rangle_b|n\rangle_c+\cdots+\psi_{d+1}(t)|d\rangle_b|n-d\rangle_c,
\end{equation}
with density matrix $\rho(t)=|\psi(t)\rangle\langle\psi(t)|$. 
Tracing out the charger degrees of freedom yields the reduced battery state $\rho_b(t)=\tr_c[\rho(t)]$.

The stored energy at time $t$ is defined as
\begin{equation}
E(t)=\tr\!\left[H_b\rho_b(t)\right]-\tr\!\left[H_b\rho_b(0)\right],
\end{equation}
which measures the energy transferred to the battery up to time $t$.
To characterize the charging performance, we consider the instantaneous charging power $P(t)=dE(t)/dt$ and the associated average charging power
\begin{equation}\label{avg_charging_power}
\bar P(\tau)=\frac{1}{\tau}\int_0^\tau P(t)\,dt=\frac{E(\tau)}{\tau},
\end{equation}
where $\tau$ is the charging time at which the stored energy reaches its maximum,
\begin{equation}
E(\tau)\equiv \max_t E(t).
\end{equation}
In the following, we focus on the charging performance at this optimal time, characterized by the average charging power $\bar P(\tau)$ and the corresponding stored energy $E(\tau)$.

\section{Structural Indicators for Charging Power and Stored Energy}\label{sec:bound}

For the initial state~\eqref{initial_state}, the effective charging Hamiltonian~\eqref{eq:tridiag} takes a tridiagonal form, in which the off-diagonal elements govern the hopping between states with different battery excitation numbers. 
This motivates us to introduce two complementary structural indicators associated with the coupling set $\{u_j(n)\}$ in the regime $n\geq N_b$,
\begin{equation}\label{Omega_n}
\Omega(n)\equiv \max_{1\le j\le N_b} u_j(n),
\end{equation}
and
\begin{equation}\label{Lambda_n}
\Lambda(n)\equiv\prod_{j=1}^{N_b} u_j(n).
\end{equation}
As shown below, $\Omega(n)$ yields an upper bound on the average charging power, whereas $\Lambda(n)$ characterizes the earliest end-to-end charging channel and thus serves as an indicator for the buildup of stored energy.

\subsection{Indicator for the average charging power}

We first derive an upper bound associated with the average charging power. 
Within the invariant subspace~\eqref{invariant_subspace}, the battery Hamiltonian in Eq.~\eqref{total_H} can be written as
\begin{equation}
\bm{H}_b\equiv H_b|_{\mathcal H_n}
=
\omega\sum_{j=0}^{N_b}\left(j-\frac{N_b}{2}\right)|j\rangle\langle j|,
\end{equation}
where, unless otherwise stated, $|j\rangle$ denotes the basis vector $|j\rangle_b|n-j\rangle_c$ in the invariant subspace $\mathcal H_n$.

Using the Heisenberg equation, $d\rho(t)/dt=-i[\bm{H},\rho(t)]$, the instantaneous charging power defined by $P(t)=dE(t)/dt$ can be expressed as
\begin{equation}
P(t)=i\,\tr\!\big(\rho(t)[\bm{H},\bm{H}_b]\big).
\end{equation}
Since the diagonal part of the Hamiltonian $\bm{H}$ commutes with $\bm{H}_b$, only the excitation-exchange terms contribute to the commutator. 
Within $\mathcal{H}_n$, one finds
\begin{equation}
[\bm{H},\bm{H}_b]=\omega \bm{K},
\end{equation}
with
\begin{equation}\label{K}
\bm{K}\equiv \sum_{j=1}^{N_b}
u_j(n)\Big(
|j-1\rangle\langle j|-|j\rangle\langle j-1|
\Big).
\end{equation}

To bound $P(t)$, we use the H\"older inequality for Schatten norms, $|\tr(A^\dagger B)|\le \|A\|_{\mathrm{tr}}\,\|B\|_{\mathrm{op}}$,
where $\|\cdot\|_{\mathrm{tr}}$ and $\|\cdot\|_{\mathrm{op}}$ denote the trace norm and the operator norm, respectively. 
Therefore, we obtain
\begin{equation}\label{P_1}
|P(t)|
=
\big|
\tr\!\left(\rho(t)\,i[\bm{H},\bm{H}_b]\right)
\big|
\le
\|\rho(t)\|_{\mathrm{tr}}\,\|[\bm{H},\bm{H}_b]\|_{\mathrm{op}}.
\end{equation}
Since $\rho(t)$ is a density operator, one has $\|\rho(t)\|_{\mathrm{tr}}=\tr[\rho(t)]=1$, and hence
\begin{equation}\label{P_1}
|P(t)|\le \|[\bm{H},\bm{H}_b]\|_{\mathrm{op}}=\omega\|\bm{K}\|_{\mathrm{op}}.
\end{equation}

We next estimate $\|\bm{K}\|_{\mathrm{op}}$ using the row-sum and column-sum matrix norms,
\begin{equation}
\|\bm{K}\|_{\mathrm{row}}\equiv\max_i\sum_j |\bm{K}_{ij}|,
\quad
\|\bm{K}\|_{\mathrm{col}}\equiv\max_j\sum_i |\bm{K}_{ij}|.
\end{equation}
Since each row and each column of $\bm{K}$ in Eq.~\eqref{K} contains at most two nonzero entries, each with magnitude no larger than $\Omega(n)$, it follows that
\begin{equation}
\|\bm{K}\|_{\mathrm{row}} \le 2\Omega(n),
\qquad
\|\bm{K}\|_{\mathrm{col}} \le 2\Omega(n).
\end{equation}
Using the standard inequality $\|\bm{K}\|_{\mathrm{op}}\le \sqrt{\|\bm{K}\|_{\mathrm{col}}\|\bm{K}\|_{\mathrm{row}}}$,
we obtain $\|\bm{K}\|_{\mathrm{op}}\le 2\Omega(n)$.
Therefore, Eq.~\eqref{P_1} becomes
\begin{equation}
|P(t)|\le 2\omega\,\Omega(n).
\label{eq:powerbound}
\end{equation}

Integrating this inequality over the charging interval $[0,\tau]$, we obtain
\begin{equation}
E(\tau)=\int_0^\tau P(t)\,dt
\le
\int_0^\tau |P(t)|\,dt
\le
2\omega\,\Omega(n)\,\tau.
\end{equation}
Hence, using Eq.~\eqref{avg_charging_power}, the average charging power satisfies
\begin{equation}
\bar P(\tau)\le 2\omega\,\Omega(n).
\label{eq:avgpowerbound}
\end{equation}
Equation~\eqref{eq:avgpowerbound} shows that $\Omega(n)$, as the largest hopping amplitude in the effective charging Hamiltonian, provides an upper bound on the average charging power. 
Equivalently, if the stored energy reaches its maximum value at the charging time $\tau$, then
\begin{equation}\label{bound_time}
\tau\ge \frac{E(\tau)}{2\omega\,\Omega(n)}.
\end{equation}
Hence, optimizing the average charging power is reduced to maximizing $\Omega(n)$ with respect to the initial charger excitation number $n$.

\subsection{The fully charged channel and the buildup of stored energy}

We now turn to the buildup of stored energy. 
Starting from the empty battery state $|0\rangle$, the charging process must progressively populate states with larger battery excitation number, ultimately approaching the fully charged state $|N_b\rangle$. 
To quantify this process, we consider the transition amplitude
\begin{equation}
\alpha(t)\equiv\langle N_b|e^{-i\bm{H}t}|0\rangle,
\end{equation}
where the shorthand \(|j\rangle\equiv |j\rangle_b|n-j\rangle_c\) within \(\mathcal H_n\) is used throughout.

Because the charging dynamics changes the battery excitation number sequentially, any transition from $|0\rangle$ to $|N_b\rangle$ must traverse all $N_b$ off-diagonal links of the effective Hamiltonian~\eqref{eq:tridiag}. 
Consequently, in the short-time expansion of $\alpha(t)$, the first nonvanishing contribution appears at order $N_b$. 
More explicitly, \(e^{-i\bm{H}t}=\sum_{m=0}^\infty (-it)^m\bm{H}^m/m!\), and one has
\begin{equation}
\langle N_b|\bm{H}^m|0\rangle=0,
\qquad m<N_b.
\end{equation}
At order $m=N_b$, there is a unique shortest path connecting $|0\rangle$ and $|N_b\rangle$, whose weight is precisely the product of all hopping amplitudes. 
Therefore,
\begin{equation}
\alpha(t)
=
\frac{(-it)^{N_b}}{N_b!}
\prod_{j=1}^{N_b}u_j(n)
+
O(t^{N_b+2}).
\end{equation}
Using Eq.~\eqref{Lambda_n}, this becomes
$\alpha(t)
=
(-it)^{N_b}\Lambda(n)/N_b!
+
O(t^{N_b+2})$.

We define the normalized stored-energy contribution associated with the fully charged battery state as
\begin{equation}
\varepsilon(t)\equiv\langle N_b|\rho(t)|N_b\rangle=|\alpha(t)|^2 .
\end{equation}
Using the short-time expansion of the transition amplitude, we obtain
\begin{equation}
\varepsilon(t)
=
\frac{t^{2N_b}}{(N_b!)^2}\Lambda^2(n)
+
O(t^{2N_b+2}).
\label{eq:channelbound}
\end{equation}

Equation~\eqref{eq:channelbound} shows that $\Lambda(n)$ governs the strength of the earliest end-to-end charging channel connecting the empty and fully charged battery states. 
Although the total stored energy depends on the full population distribution over all battery-excitation sectors, a larger $\Lambda(n)$ enhances the short-time transfer toward highly excited battery states and thus favors more efficient energy storage. 
In this sense, $\Lambda(n)$ provides a natural structural indicator for the buildup of stored energy.

The two structural indicators introduced above characterize two complementary aspects of the same optimization problem. 
The quantity $\Omega(n)$ determines the largest hopping amplitude and yields an upper bound on the average charging power, Eq.~\eqref{eq:avgpowerbound}, whereas $\Lambda(n)$ quantifies the earliest end-to-end charging channel toward the fully charged configuration, Eq.~\eqref{eq:channelbound}. 
In the next section, we show that these two indicators are jointly optimized at the same initial charger excitation number, which reveals the emergent mirror symmetry of the optimal charging configuration.


\section{Emergent Mirror Symmetry in the Optimal Charging Configuration}\label{sec:mirror}

We now analyze the $n$ dependence of the effective hopping amplitudes $u_j(n)$ given in Eq.~\eqref{off_diagonal}, together with the two structural indicators $\Omega(n)$ and $\Lambda(n)$ defined in Eqs.~\eqref{Omega_n} and \eqref{Lambda_n}. 
Our goal is to determine the value of $n$ that jointly optimizes these two quantities.

\subsection{Common optimality of $\Lambda(n)$ and $\Omega(n)$}

We first consider the channel indicator $\Lambda(n)$. 
From Eq.~\eqref{off_diagonal}, one has
\begin{equation}
\Lambda^2(n)
=
A^{2N_b}
\prod_{j=1}^{N_b} j(N_b-j+1)\, G(n),
\end{equation}
where
\begin{equation}
G(n)\equiv \prod_{j=1}^{N_b}(N_c-n+j)(n-j+1).
\end{equation}
Since the prefactor $\prod_{j=1}^{N_b} j(N_b-j+1)$ is independent of $n$, maximizing $\Lambda(n)$ is equivalent to maximizing $G(n)$.

The monotonicity of $G(n)$ is determined by the ratio
\begin{equation}
\frac{G(n+1)}{G(n)}
=
\frac{(N_c-n)(n+1)}{(N_c-n+N_b)(n-N_b+1)}.
\label{eq:G_ratio}
\end{equation}
Therefore, $G(n+1)\ge G(n)$ holds if and only if
\begin{equation}
(N_c-n)(n+1)\ge (N_c-n+N_b)(n-N_b+1),
\end{equation}
which is equivalent to
\begin{equation}
N_b\bigl(N_b+N_c-2n-1\bigr)\ge 0.
\end{equation}
Hence, $G(n)$ is increasing for $n<(N_b+N_c-1)/2$ and decreasing for $n>(N_b+N_c-1)/2$. 
This proves that $\Lambda(n)$ attains its maximum at the integer nearest to the center of the admissible interval. 
We therefore define
\begin{equation}
n_*\equiv \left\lfloor \frac{N_b+N_c}{2}\right\rfloor
\label{eq:nstar}
\end{equation}
as the excitation number that maximizes $\Lambda(n)$.

We next turn to the indicator $\Omega(n)$, namely the largest hopping amplitude in the effective charging Hamiltonian. 
From Eq.~\eqref{off_diagonal}, one has
\begin{equation}
u_j^2(n)=A^2\,j(N_b-j+1)(N_c-n+j)(n-j+1).
\label{eq:uj_square}
\end{equation}
To estimate the maximal value of $\Omega(n)$, we bound the two factors on the right-hand side separately. 
For the battery-dependent factor in Eq.~\eqref{eq:uj_square}, the arithmetic--geometric mean inequality gives
\begin{equation}
j(N_b-j+1)\le \left(\frac{N_b+1}{2}\right)^2,
\label{eq:battery_factor_bound}
\end{equation}
while for the charger-dependent factor in Eq.~\eqref{eq:uj_square} one has
\begin{equation}
(N_c-n+j)(n-j+1)\le \left(\frac{N_c+1}{2}\right)^2.
\label{eq:charger_factor_bound}
\end{equation}
Combining Eqs.~\eqref{eq:battery_factor_bound} and \eqref{eq:charger_factor_bound}, we obtain
\begin{equation}
u_j^2(n)\le
A^2\left(\frac{N_b+1}{2}\right)^2
\left(\frac{N_c+1}{2}\right)^2,
\end{equation}
and hence
\begin{equation}\label{eq:Omega_bound}
\Omega(n)\le A\,\frac{N_b+1}{2}\,\frac{N_c+1}{2}.
\end{equation}

The first inequality in Eq.~\eqref{eq:battery_factor_bound} is saturated when $j=(N_b+1)/2$, while the second inequality in Eq.~\eqref{eq:charger_factor_bound} is saturated when $N_c-n+j=n-j+1$. 
Combining these two saturation conditions, one obtains $n=(N_b+N_c)/2$. 
Since $n$ must be an integer, the corresponding optimal excitation number is represented by $n_*$ defined in Eq.~\eqref{eq:nstar}. 
We thus conclude that the two structural indicators $\Lambda(n)$ and $\Omega(n)$ are jointly optimized at the same excitation number $n=n_*$.

\begin{figure}[t]
\centering
\includegraphics[width=1\linewidth]{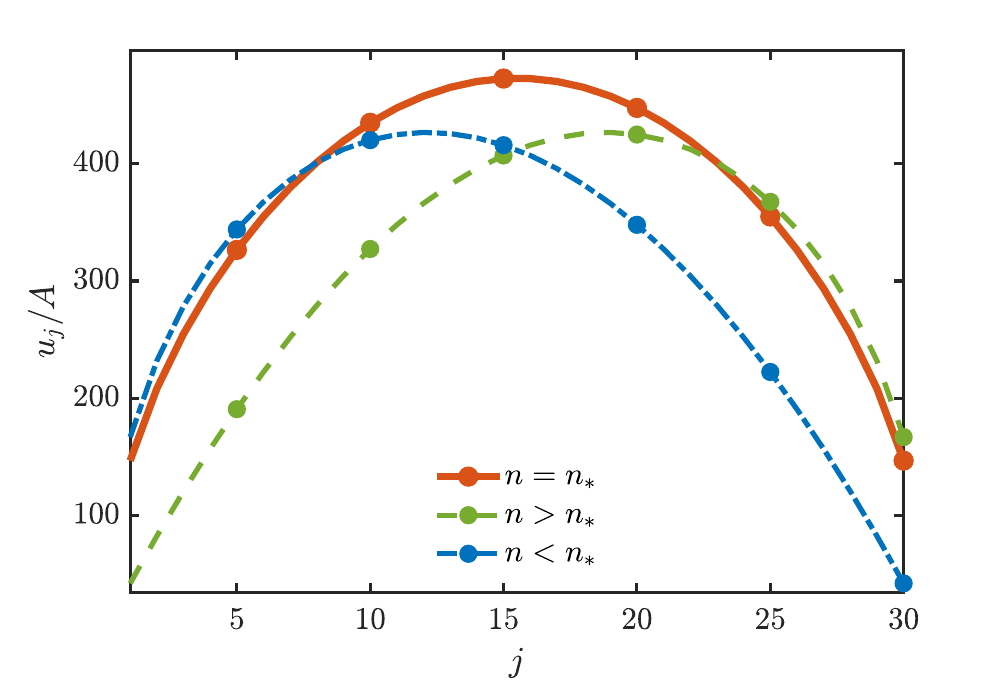}
\caption{
Spatial profile of the effective hopping amplitudes \(u_j\) for different initial charger excitation numbers \(n\). 
At the optimal point \(n=n_*\), the profile is exactly mirror symmetric with respect to the chain center, whereas away from this point it becomes asymmetric. 
Here \(n_*=(N_b+N_c)/2\) for the even case considered. 
The parameters are \(N_b=30\), \(N_c=90\), and \(A=1\).
}
\label{fig:symmetry}
\end{figure}

\begin{figure}[t]
\centering
\includegraphics[width=1\linewidth]{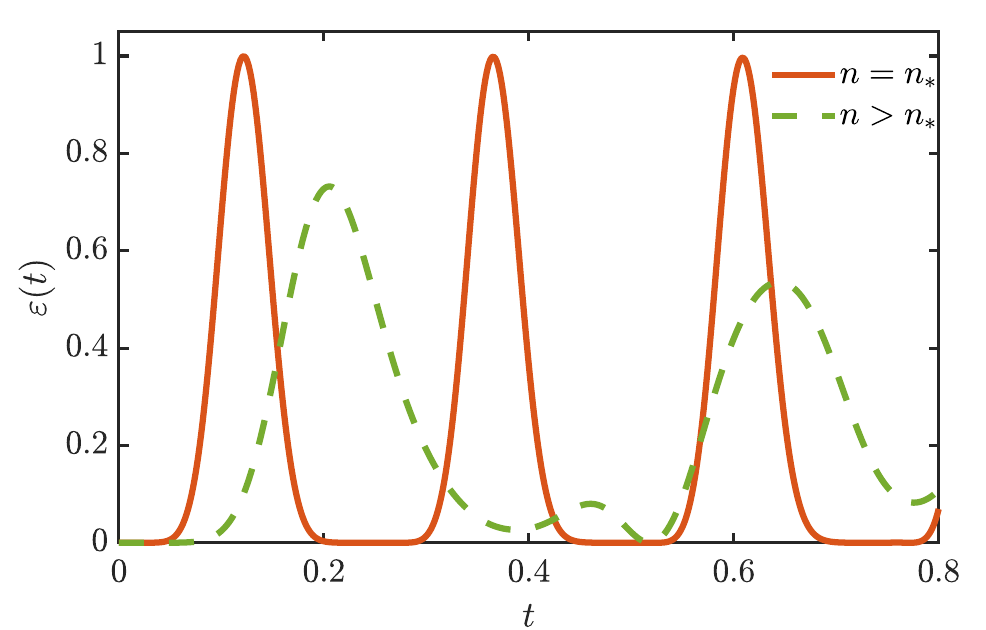}
\caption{
Time evolution of the normalized stored-energy contribution associated with the fully charged battery state, 
\(\varepsilon(t)=\langle N_b|\rho(t)|N_b\rangle\), 
for different initial charger excitation numbers \(n\). 
At the optimal point \(n=n_*\), the transfer to the fully charged state is the most efficient, leading to the largest and fastest buildup of \(\varepsilon(t)\). 
Here \(n_*=(N_b+N_c)/2\) for the even case considered. 
The parameters are \(N_b=5\), \(N_c=25\), and \(A=1\).
%
}
\label{fig:dynamics}
\end{figure}

\subsection{Mirror symmetry at the common optimal point}

We now examine the structure of the effective charging Hamiltonian at the common optimal point \(n=n_*\). 
For clarity, we take the even case as a representative example, where \(N_b+N_c\) is even and hence \(n_*=(N_b+N_c)/2\).
For the tridiagonal Hamiltonian in Eq.~\eqref{eq:tridiag}, the diagonal entries are constant and therefore automatically mirror symmetric. 
Mirror symmetry is thus determined by the hopping amplitudes, which must be invariant under reflection about the center of the charging pathway, i.e.,
\begin{equation}
u_j(n)=u_{N_b-j+1}(n),
\qquad j=1,\dots,N_b.
\label{eq:mirror_def}
\end{equation}
Equivalently, the left and right halves of the hopping profile are exactly matched with respect to the center of the effective chain.

This symmetry can be seen directly from the explicit form of the hopping amplitudes in Eq.~\eqref{off_diagonal}. 
Under the transformation
\begin{equation}
n\mapsto \widetilde n\equiv N_b+N_c-n,
\label{eq:mirror_map}
\end{equation}
one finds
\begin{equation}
u_j(\widetilde n)=u_{N_b-j+1}(n),
\qquad j=1,\dots,N_b.
\label{eq:mirror_covariance}
\end{equation}
Thus, the reflection $n\mapsto \widetilde n$ exchanges the hopping amplitudes on the two sides of the effective charging pathway.

At the common optimal point $n=n_*=(N_b+N_c)/2$, one has $\widetilde n_*=n_*$. 
Equation~\eqref{eq:mirror_covariance} then reduces to
\begin{equation}
u_j(n_*)=u_{N_b-j+1}(n_*),
\qquad j=1,\dots,N_b,
\end{equation}
which is precisely the mirror-symmetry condition~\eqref{eq:mirror_def}. 
Therefore, the effective charging Hamiltonian becomes exactly mirror symmetric at the point where the two indicators are jointly optimized.

Figure~\ref{fig:symmetry} provides a direct visualization of the hopping profiles for different initial charger excitation numbers. 
At the common optimal point $n=n_*$, the hopping amplitudes are exactly symmetric with respect to the center of the effective chain. 
By contrast, when $n$ deviates from this value, the profile becomes visibly asymmetric, indicating that the charging pathway is biased toward one side. 
The figure therefore shows that mirror symmetry is the structural signature of the optimal effective Hamiltonian.

\begin{figure}[t]
\centering
\includegraphics[width=0.9\linewidth]{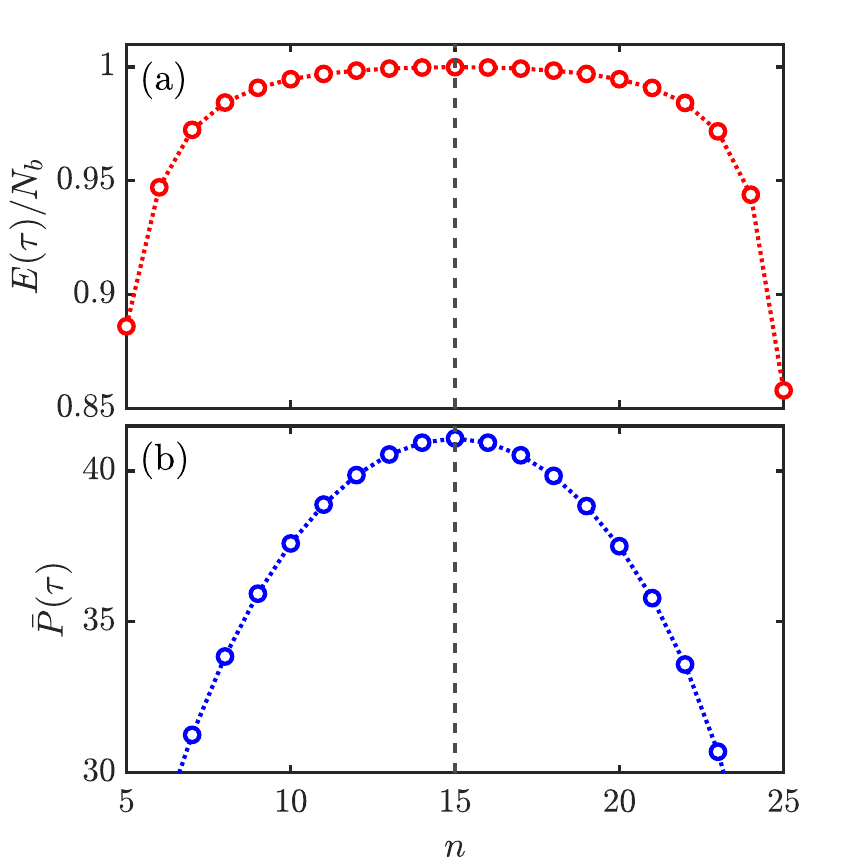}
\caption{
Charging performance as a function of the initial charger excitation number \(n\). 
The maximum stored energy per battery cell \( E(\tau)/N_b\) in (a) and the corresponding average charging power \(\bar P(\tau)\) in (b) are both maximized at the common optimal point \(n=n_*\), marked by the vertical dashed line. 
Here \(n_*=(N_b+N_c)/2\) for the even case considered. 
The parameters are \(N_b=5\), \(N_c=25\), and \(A=1\).
}
\label{fig:optimal}
\end{figure}

\subsection{Mirror symmetry and its dynamical consequence}

The dynamical consequence of mirror symmetry is illustrated in Fig.~\ref{fig:dynamics}. 
At the mirror-symmetric point $n=n_*$, the population transfer to the fully charged battery state is the most efficient, with $\varepsilon(t)=\langle N_b|\rho(t)|N_b\rangle$ developing nearly unit peaks. 
Away from this point, the transfer becomes less efficient and the peak population is significantly reduced. 
This shows that the emergent mirror symmetry is not merely a static property of the effective Hamiltonian, but directly enhances the charging dynamics.

The same conclusion is confirmed at the level of physical charging observables. 
As shown in Fig.~\ref{fig:optimal}, both the maximum stored energy $E(\tau)$ and the corresponding average charging power $\bar P(\tau)$ are maximized at $n=n_*$. 
Therefore, the symmetry-selected point identified from the effective hopping profile is also the operating point that optimizes the physically relevant performance of the quantum battery. 
Taken together with Figs.~\ref{fig:symmetry},~\ref{fig:dynamics} and \ref{fig:optimal}, this establishes a consistent picture: the optimization of the charging indicators selects a mirror-symmetric hopping profile, which in turn leads to more efficient charging dynamics and optimal battery performance.

The usefulness of mirror symmetry is also physically transparent. 
When the hopping profile is mirror symmetric, the charging pathway from the empty state to the fully charged state is balanced about its center, which favors coherent end-to-end transfer. 
More generally, mirror symmetry is a key structural ingredient for perfect state transfer in one-dimensional hopping problems with an appropriate spectral condition~\cite{christandl2004perfect,moosavi2026perfect,albanese2004mirror,karbach2005spin,vinet2012construct}. 
This explains why the symmetry-selected optimal point $n=n_*$ is favorable for efficient charging and energy storage.

A further implication of the symmetry-selected optimal point is the scaling of the charging time with the charger size. 
According to Eq.~\eqref{eq:Omega_bound}, at the symmetry-selected optimal point with an oversized charger, $N_c>N_b$, one has
\begin{equation}
\Omega_{\mathrm{opt}}
=
A\left(\frac{N_b+1}{2}\right)\left(\frac{N_c+1}{2}\right)
\approx
\frac{A}{4}N_bN_c.
\end{equation}
Substituting this result into Eq.~\eqref{bound_time}, we obtain
\begin{equation}
\tau_{\mathrm{opt}}
\gtrsim
\frac{2E(\tau)}{\omega A N_b N_c}.
\end{equation}
Therefore, for fixed $N_b$ and large $N_c$, the optimal charging time obeys the scaling law
\begin{equation}
\tau_{\mathrm{opt}}\propto \frac{1}{N_c}.
\end{equation}

\begin{figure}[t]
\centering
\includegraphics[width=1\linewidth]{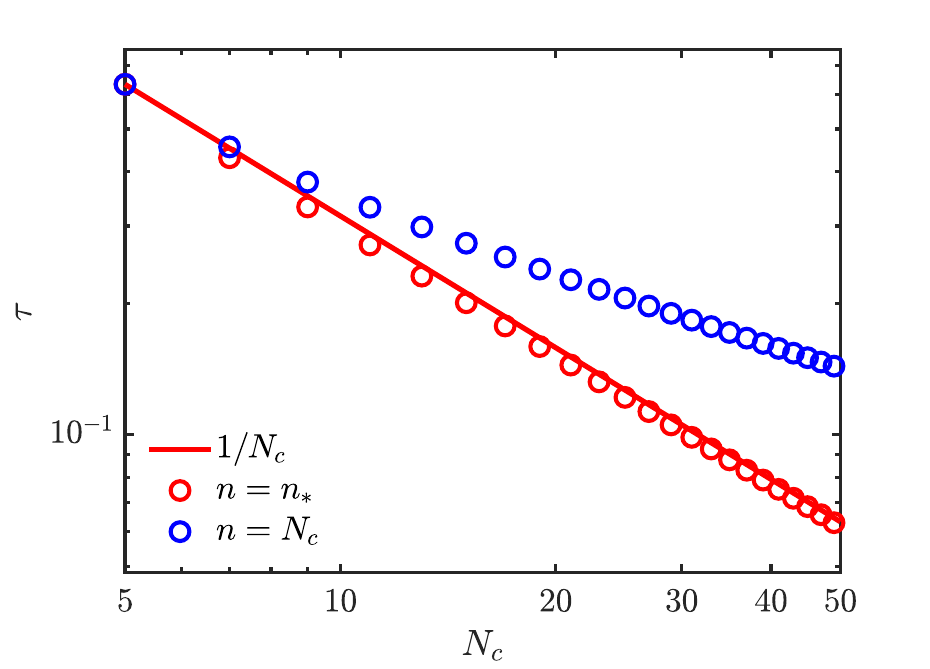}
\caption{
Scaling of the charging time $\tau$ with the charger size $N_c$. 
Red circles denote the numerical results for the symmetry-selected optimal protocol $n=n_*$, blue circles denote those for the fully excited charger baseline $n=N_c$, and the red solid line is the $1/N_c$ scaling guide. 
The optimal protocol follows $\tau\propto 1/N_c$, whereas the fully excited charger baseline does not. 
Here $n_*=(N_b+N_c)/2$ for the even case considered. 
The parameters are $N_b=5$ and $A=1$.
}
\label{fig:scaling}
\end{figure}

Figure~\ref{fig:scaling} confirms this prediction by comparing the symmetry-selected protocol with the fully polarized benchmark. 
The red circles, corresponding to the optimal Dicke-state protocol at $n=n_*$, are in excellent agreement with the $1/N_c$ reference scaling marked by the red solid line. 
In contrast, the blue circles, corresponding to the fully polarized charger scheme $n=N_c$, decrease much more slowly and clearly deviate from the $1/N_c$ behavior. 
This comparison demonstrates that the reduction of the charging time is not a trivial consequence of enlarging the charger, but originates from the structure of the symmetry-selected optimal initial state.


\begin{figure*}[t]
\centering
\includegraphics[width=1\linewidth]{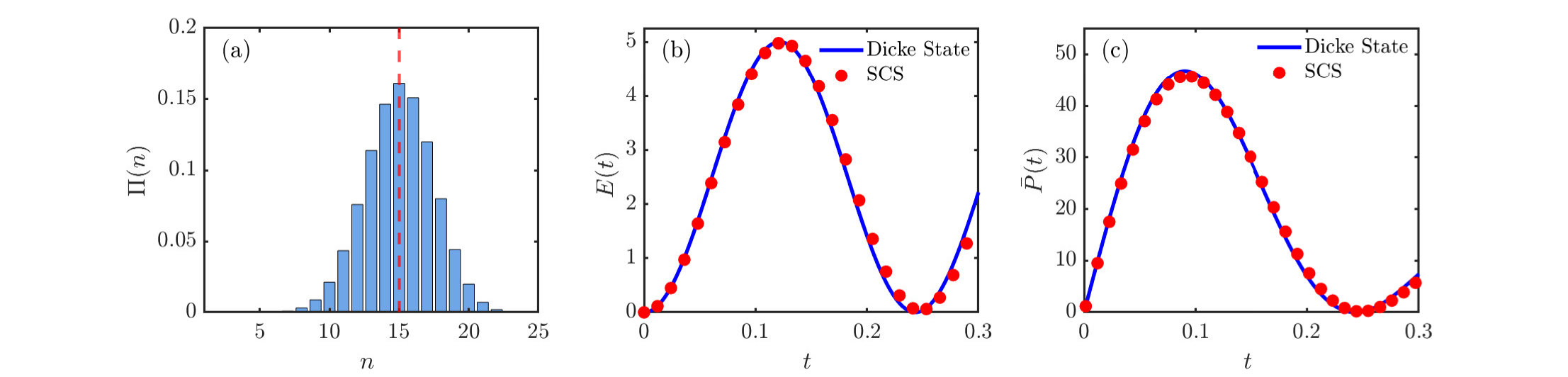}
\caption{
Spin-coherent-state approximation to the symmetry-selected Dicke protocol. 
(a) Excitation-number distribution $\Pi(n)$ of the optimal spin coherent state, with the red dashed line marking the symmetry-selected point $n=n_*$. 
(b) Stored energy $E(t)$ and (c) average charging power $\bar P(t)$ for the optimal Dicke state (blue solid lines) and the corresponding spin coherent state (red circles) prepared with $\theta_{\mathrm{opt}}=\arccos(N_b/N_c)$. 
The close agreement shows that the spin coherent state captures the main features of the optimal charging dynamics. 
The parameters are $N_b=5$, $N_c=25$, and $A=1$.
%
}
\label{fig:scs}
\end{figure*}

\section{Charging Dynamics from Product Initial States}\label{sec:simulate}

The optimal charging protocol identified above is associated with the Dicke sector $n=n_*$, where the effective hopping profile becomes mirror symmetric. 
It is therefore natural to ask to what extent the corresponding dynamics can be approximated by simpler initial states of product form. 
Such product states are also more accessible than highly entangled Dicke states from the viewpoint of state preparation. 
To address this question, we consider a charger prepared in a spin coherent state
\begin{equation}
|\theta,\phi\rangle
=
\bigotimes_{k=1}^{N_c}
\left(
\cos\frac{\theta}{2}\,|\uparrow\rangle_k
+
e^{i\phi}\sin\frac{\theta}{2}\,|\downarrow\rangle_k
\right).
\end{equation}
Since the excitation-number distribution depends only on $\theta$, the phase $\phi$ does not affect the present analysis and will be set to zero in the following. 
Unlike the Dicke state, this state is separable, but it can be expanded in the Dicke basis as
\begin{equation}
|\theta,0\rangle=\sum_{n=0}^{N_c} C_n(\theta)\,|n\rangle_c,
\end{equation}
with excitation-number distribution
\begin{equation}
\Pi(n)=|C_n(\theta)|^2=
\binom{N_c}{n}
\left(\cos^2\frac{\theta}{2}\right)^n
\left(\sin^2\frac{\theta}{2}\right)^{N_c-n}.
\end{equation}
For large $N_c$, the distribution $\Pi(n)$ is sharply peaked around the mean $\mu=N_c\cos^2(\theta/2)$ with variance $\sigma^2=N_c\cos^2(\theta/2)\sin^2(\theta/2)$. 
Thus, although the spin coherent state is not confined to a single Dicke sector, its dynamics is dominated by the sectors carrying the largest statistical weight.

To approximate the symmetry-selected optimal Dicke state, one should align the center of the excitation-number distribution with the optimal sector, namely $\mu=n_*$. 
Using $\mu=N_c\cos^2(\theta/2)$, this condition determines the optimal polar angle as
\begin{equation}
\theta_{\mathrm{opt}}=\arccos\left(\frac{N_b}{N_c}\right).
\end{equation}
At the same time, the relative width scales as $\sigma/\mu\sim 1/\sqrt{N_c}$, so that, for sufficiently large $N_c$, the dominant weight of the spin coherent state remains concentrated near the symmetry-selected sector. 
In this sense, the spin coherent state provides a product-state approximation to the optimal Dicke-state protocol.

The physical picture is straightforward. 
The exact mirror-symmetric hopping profile is realized only in the Dicke sector $n=n_*$. 
However, if the initial product state has an excitation-number distribution centered at $n_*$ and sufficiently narrow around it, then the resulting dynamics is dominated by sectors whose effective hopping profiles remain close to the mirror-symmetric one. 
As a consequence, the charging evolution can still inherit the main features of the optimal protocol, including efficient energy transfer and enhanced charging power.

Figure~\ref{fig:scs} illustrates this approximation explicitly. 
Panel (a) shows that the excitation-number distribution of the optimally chosen spin coherent state is centered at the symmetry-selected value. 
Panels (b) and (c) compare the stored energy and the average charging power with those obtained from the optimal Dicke-state protocol. 
The close agreement demonstrates that aligning the excitation-number distribution is sufficient to reproduce the main dynamical features of the ideal symmetry-selected charging process. 
Therefore, near-optimal charging does not require exact preparation of the Dicke state, but can already be captured by an appropriate product initial state, which is simpler from the viewpoint of state preparation.

\section{Conclusions}\label{sec:conclusion}

In this work, we studied the optimization of a quantum battery based on the central-spin model. 
By restricting the dynamics to the invariant excitation subspace, we identified two complementary structural indicators associated with the effective hopping profile: the largest hopping amplitude $\Omega(n)$ and the end-to-end channel indicator $\Lambda(n)$. 
The former yields an upper bound on the average charging power, while the latter characterizes the earliest buildup of the fully charged channel and thus serves as an indicator for the buildup of stored energy. 
We showed that both $\Omega(n)$ and $\Lambda(n)$ are optimized at the same excitation number $n=n_*$, which selects a distinguished Dicke sector of the model. 
At this common optimal point, the effective hopping profile becomes mirror symmetric, \(u_j(n_*)=u_{N_b-j+1}(n_*)\), showing that mirror symmetry is not imposed externally, but instead emerges directly from the optimization problem itself. 
This symmetry-selected structure has clear dynamical consequences: it favors efficient population transfer from the empty battery state to the fully charged state, simultaneously optimizes the stored energy at the optimal charging time and the corresponding average charging power, and leads to a favorable charging-time scaling for oversized chargers.

We further showed that the optimal Dicke-state protocol can be closely approximated by product initial states. 
In particular, spin coherent states reproduce the main features of the optimal charging dynamics when their excitation-number distribution is centered at the symmetry-selected point and remains sufficiently narrow around it. 
This demonstrates that the essential features of the optimal charging dynamics do not rely on exact preparation of the Dicke state, but can already be captured by appropriate product initial states. 
Our results establish a direct connection between charging performance, optimal-state structure, and emergent symmetry in the central-spin quantum battery, and suggest symmetry as a useful organizing principle for understanding and designing efficient charging protocols in interacting many-body quantum systems.


\begin{acknowledgments}
This work was supported by the NSFC (Grants No.12275215, No.12305028, and No.12247103), and the Youth Innovation Team of Shaanxi Universities. KZ is supported by the China Postdoctoral Science Foundation under Grant Number 2025M773421, Shaanxi Province Postdoctoral Science Foundation under Grant Number 2025BSHYDZZ017, and Scientific Research Program Funded by Education Department of Shaanxi Provincial Government (Program No.24JP186).
HLS was supported by the European Commission through the
H2020 QuantERA ERA-NET Cofund in Quantum Technologies project ``MENTA'' and received
funding under Horizon Europe programme HORIZONCL4-2022-QUANTUM-02-SGA via the
project 101113690 (PASQuanS2.1).
\end{acknowledgments}

\bibliography{ref}

\end{document}